\documentclass[preprintnumbers,prd,twocolumn,tightenlines,floatfix,showpacs,amssymb,nofootinbib]{revtex4}

\usepackage{pictex}
\usepackage[dvips]{graphicx}
\usepackage{amsmath}

\usepackage{amsmath}
\usepackage{amsfonts}
\usepackage{amsthm}

\newcommand{\Dslash}{\!\!\!\!/\,}

\newcommand{\nn}{\nonumber}
\newcommand{\half}{\frac{1}{2}}
\newcommand{\quarter}{\frac{1}{4}}
\newcommand{\eqn}[1]{ \begin{equation} #1 \end{equation} }

\newcommand{\link}{
\setlength{\unitlength}{14pt}
\begin{picture}(1,1)(0,0)
\linethickness{0.25pt}
\put(0,0){\circle*{0.15}}
\put(0,0){\vector(1,0){1}}
\put(1,0){\circle{0.14}}
\end{picture}}
\newcommand{\stapleup}{
\setlength{\unitlength}{14pt}
\begin{picture}(1,1)(0,0)
\linethickness{0.25pt}
\put(0,0){\circle*{0.15}}
\put(0,0){\vector(0,1){1}}
\put(0,1){\vector(1,0){1}}
\put(1,1){\vector(0,-1){1}}
\put(1,0){\circle{0.14}}
\end{picture}}
\newcommand{\stapledown}{
\raisebox{-14pt}{
\setlength{\unitlength}{14pt}
\begin{picture}(1,1)(0,-1)
\linethickness{0.25pt}
\put(0,0){\circle*{0.15}}
\put(0,0){\vector(0,-1){1}}
\put(0,-1){\vector(1,0){1}}
\put(1,-1){\vector(0,1){1}}
\put(1,0){\circle{0.14}}
\end{picture}}}

\begin{document}
\preprint{ADP-07-07/T647}


\title{Scaling analysis of FLIC fermion actions}

\author{Waseem Kamleh}
\author{Ben Lasscock}
\author{Derek B. Leinweber}
\author{Anthony G. Williams}
\affiliation{Special Research Centre for the Subatomic Structure of Matter and Department of Physics, University of Adelaide 5005, Australia. }

\begin{abstract}
The Fat Link Irrelevant Clover (FLIC) fermion action is a variant of the $O(a)$-improved Wilson action where the irrelevant operators are constructed using smeared links. While the use of such smearing allows for the use of highly improved definitions of the field strength tensor $F_{\mu\nu},$ we show that the standard 1-loop clover term with a mean field improved coefficient $c_{\rm sw}$ is sufficient to remove the $O(a)$ errors, avoiding the need for non-perturbative tuning. This result enables efficient dynamical simulations in QCD with the FLIC fermion action
\end{abstract}

\pacs{
11.15.Ha,  
12.38.Gc  
}

\maketitle

\section{Introduction}

The fat-link irrelevant clover (FLIC) fermion action
\cite{zanotti-hadron} is an efficient \cite{kamleh-spin} Wilson-style
nearest-neighbour lattice fermion action which incorporates both the
thin gauge-field links of the Markov chain and fat links -- links
created via APE \cite{ape-one,ape-two,derek-smooth,ape-MIT}, HYP
\cite{hyp-smear} or stout-link \cite{stout-links} smearing.  Through
the use of fat links in the irrelevant operators of the action, one
achieves significant improvement in the chiral properties of the action
reflected in a narrowing of the distribution of the critical Wilson
mass \cite{flic-impchiral}.  One also by-passes the fine-tuning
problem typically encountered in ${\cal O}(a)$ improvement, as the use
of fat links in both the irrelevant Wilson and clover terms suppresses
the otherwise large renormalizations of the improvement coefficients.
At the same time, short-distance physics is preserved completely in
the action as the relevant operators are constructed with thin links.

Previous work \cite{james-scale} established the good scaling
properties of the Fat Link Irrelevant Clover (FLIC) fermion action
when a highly improved definition of the lattice field strength tensor
$F_{\mu\nu}$ is used in the clover term. In this work we demonstrate
that the use of the standard 1-loop definition of $F_{\mu\nu}$ with
fat links in the clover term is sufficient to provide $O(a^2)$ scaling
for FLIC fermions. The 1-loop variant has the advantage of maintaining
a simple force term when performing the molecular dynamics portion of
a Hybrid Monte Carlo algorithm to generate dynamical configurations.

In Sec.~\ref{FLICfermions} we highlight the essential features of the
FLIC action with a particular emphasis on the various lattice field
strength tensors used in the simulations.  In addition the
$SU(3)$-projection method used to create the fat links is outlined.
In Sec.~\ref{scaleDetermination} we describe the methods used to
obtain an accurate scale determination on each lattice considered.
Simulation parameters and scaling results are presented in
Sec.~\ref{scalingResults} while correlation function properties are
examined in Sec.~\ref{correlationFunctions}.  Conclusions are
summarized in Sec.~\ref{conclusions}.

\section{FLIC Fermions}
\label{FLICfermions}

The FLIC fermion action \cite{zanotti-hadron} is a variant of the
clover action where the irrelevant operators are constructed using
smeared links~\cite{ape-one,ape-two}, and mean field
improvement~\cite{lepage-mfi} is performed. The key point is that
short-distance physics is suppressed in the irrelevant operators.  This
allows an effective mean-field improved calculation of the clover
coefficient, required to match the Wilson and clover terms such that
$O(a)$ errors are eliminated \cite{james-scale}.  Further, the improved
chiral properties of FLIC fermion action allow efficient access to the
light quark regime \cite{flic-impchiral}. 

The FLIC operator is given by
\eqn{D_{\rm FLIC} = \nabla\Dslash_{\rm mfi} + \frac{1}{2}(\Delta^{\rm
fl}_{\rm mfi} - \frac{1}{2}\sigma\cdot F^{\rm fl}_{\rm mfi}) - m \, ,}
where the presence of fat (or smeared) links and/or mean field
improvement has been indicated by the super- and subscripts. The mean
field improved lattice gauge covariant derivative is defined by
\eqn{\nabla\Dslash_{\rm mfi} = \sum_\mu
\frac{1}{2u_0}\gamma_\mu\left(U_\mu(x)\, \delta_{x+\hat{\mu},y} -
U_\mu^\dagger(x-\hat{\mu})\, \delta_{x-\hat{\mu},y}\right),} 
and likewise the (smeared link) lattice Laplacian is such that
\eqn{\Delta^{\rm fl}_{\rm mfi} = \sum_\mu 2 - \frac{1}{u^{\rm
fl}_0}\left(U^{\rm fl}_\mu(x)\, \delta_{x+\hat{\mu},y} + U_\mu^{{\rm
fl}\dagger}(x-\hat{\mu})\, \delta_{x-\hat{\mu},y}\right).}

We choose $\sigma_{\mu\nu}=\frac{i}{2}[\gamma_\mu,\gamma_\nu].$ For
the clover term, one usually selects a standard one-loop $F_{\mu\nu},$
\begin{align}
F_{\mu\nu}(x) &= -\frac{i}{2}(C_{\mu\nu}(x) - C^\dagger_{\mu\nu}(x)), \\
\nn C_{\mu\nu}(x) &= \quarter(U_{\mu,\nu}(x) + U_{-\nu,\mu}(x) \\
&\qquad + U_{\nu,-\mu}(x) + U_{-\mu,-\nu}(x)),
\end{align}
where $U_{\mu,\nu}(x) = U_\mu(x)\, U_\nu(x+\hat{\mu})\,
U^\dagger_\mu(x+\hat{\nu})\, U^\dagger_\nu(x)$ is the elementary
plaquette in the $\mu,\nu$ plane.  However, with the use of fat links,
one is also able to choose highly improved definitions of $F_{\mu\nu}$
\cite{sundance-fmunu}.  Let $C^{m\times n}_{\mu\nu}(x)$ correspond to
the sum of the four $m\times n$ loops at the point $x$ in the clover
formation, and then define
\eqn{ F^{m\times n}_{\mu\nu}(x) = -\frac{i}{2}(C^{m\times
    n}_{\mu\nu}(x) - C^{\dagger m\times n}_{\mu\nu}(x)).} 
We can construct a 2-loop field strength tensor which is free of
$O(a^2)$ errors,
\eqn{ F^{\rm 2L}_{\mu\nu} = \frac{5}{3} F^{1\times 1}_{\mu\nu} -\frac{1}{6} (F^{1\times 2}_{\mu\nu} + F^{2\times 1}_{\mu\nu}),}
or a 3-loop version which is free of $O(a^4)$ errors,
\eqn{ F^{\rm 3L}_{\mu\nu} = \frac{3}{2} F^{1\times 1}_{\mu\nu} -\frac{3}{20} F^{2\times 2}_{\mu\nu} + \frac{1}{90} F^{3\times 3}_{\mu\nu}.}

The smeared links in the FLIC action can be equally well constructed
from standard APE smeared links, or the more novel stout link
method \cite{stout-links}. As the smeared links only appear in the
irrelevant operators, the physics of the action are essentially
independent of the choice of smearing method. The only requirement is
that sufficient smearing is done such that the mean field improvement
becomes an effective means of estimating the clover coefficient
$c_{\rm sw}.$ We typically find that four sweeps of APE smearing at
$\alpha=0.7$ or four sweeps of stout smearing at $\rho = 0.1$ to be
sufficient for lattices with a spacing between 0.1 and 0.165 fm.

In this work we use APE smeared links $U^{\rm fl}_\mu(x)$ constructed
from $U_\mu(x)$ by performing $4$ smearing sweeps, where in each sweep
we first perform an APE blocking step (at $\alpha=0.7$),
\begin{equation}
V^{(j)}_\mu(x) = (1-\alpha)\ \link + \frac{\alpha}{6} \sum_{\nu \ne \mu}\ \stapleup\ + \stapledown\ , 
\end{equation}
followed by a projection back into $SU(3), U^{(j)}_\mu(x) = {\mathcal
P}(V^{(j)}_\mu(x)).$ We follow the ``unit-circle'' projection method
given in \cite{kamleh-hmc}, which allows for dynamical
simulations.  The projection is defined by first performing a
projection into $U(3)$
\begin{equation}
U'(V) = V [V^\dagger V]^{-\half},
\end{equation}
followed by projection into $SU(3)$
\begin{equation}
{\mathcal P}(V) = \frac{1}{\sqrt[3]{\det U'(V)}} \, U'(V) \, .
\end{equation}
It should be noted that the principal value of the cube root (being
that with the largest real part) is the appropriate branch of the cube
root function to choose. As noted in \cite{kamleh-hmc} this choice
provides the mean link which is closest to unity.

Mean field improvement is performed by making the replacements
\eqn{ U_\mu(x) \to \frac{U_\mu(x)}{u_0},\quad U^{\rm fl}_\mu(x) \to
  \frac{U^{\rm fl}_\mu(x)}{u^{\rm fl}_0}, } 
where $u_0$ and $u_0^{\rm fl}$ are the mean links for the standard and
fattened links. We calculate the mean link via the fourth root of the
average plaquette
\eqn{ u_0 = \left \langle {\rm \frac{1}{3}\, Re\, Tr }\,  U_{\mu\nu}(x)
  \right \rangle_{x,\, \mu<\nu}^{\frac{1}{4}} \, . }

\section{Scale determination}
\label{scaleDetermination}

The scale is determined using a 4-parameter ansatz
\eqn{\label{eq:4fit} V({\rm\bf r}) = V_0 + \sigma\, r -
  e\left[\frac{1}{\rm\bf r}\right]+l\left(\left[\frac{1}{\rm\bf
      r}\right]-\frac{1}{r}\right)} 
as in Ref.~\cite{accurate-scale}. The tree-level lattice Coulomb term
used in the ansatz is given by 
\eqn{ \left[\frac{1}{\rm\bf r}\right] = 4\pi \int \frac{d^3{\rm\bf
      k}}{2\pi^3}\, \cos ({\rm\bf k\cdot r})\, D_{00}(0,{\rm\bf k}). } 
Here $D_{00}(0,{\rm\bf k})$ comes from the tree-level gluon propagator for the appropriate gluon action. For the Wilson gluon action, we have at tree-level,
\eqn{ D^{-1}_{00}(0,{\rm\bf k}) = 4\sum_{\mu=1}^3 \sin^2 \frac{k_\mu}{2},}
where on a lattice with extents $L_\mu$ the allowed momenta are 
\eqn{ k_\mu = \frac{2\pi n_\mu}{L_\mu},\quad -\frac{L_\mu}{2} < n_\mu
  \le \frac{L_\mu}{2} \, .}
For the L\"{u}scher-Weisz gluon action, we have at tree-level,
\eqn{ D^{-1}_{00}(0,{\rm\bf k}) = 4\sum_\mu \left( \sin^2 \frac{k_\mu}{2} + \frac{1}{3}\sin^4 \frac{k_\mu}{2}\right).}

The lattice Coulomb term is constructed by calculating on large
lattice volumes and then extrapolating to infinite volume.
Explicitly, we choose $L=128$ and $L=256$ and calculate
$\left[\frac{1}{\rm\bf r}\right]_L$ for an $L^3$ spatial volume.  On a
finite volume, the Coulomb term takes the form \cite{urs-private}
\eqn{\frac{1}{r} + \frac{1}{L-r} = \frac{1}{r} + \frac{1}{L} + O\left(\frac{r}{L^2}\right).}
In order to calculate the infinite volume tree-level lattice Coulomb
term $\left[\frac{1}{\rm\bf r}\right]$, we extrapolate
$\left[\frac{1}{\rm\bf r}\right]_L$ linearly in $\frac{1}{L}$ to
$\frac{1}{L}=0$. 

\begin{figure}[t]
\includegraphics[angle=90,width=\hsize]{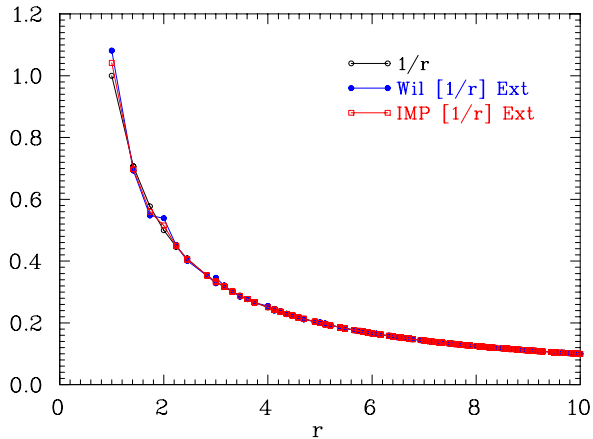}
\caption{The (infinite volume) tree-level lattice Coulomb term for the
  Wilson and L\"{u}scher-Weisz (IMP) gauge action.  } 
\label{fig:coulomb}
\end{figure}

The tree-level lattice Coulomb term $\left[\frac{1}{\rm\bf r}\right]$
for the Wilson and L\"{u}scher-Weisz gauge action is shown in
Fig~\ref{fig:coulomb}.  The important finite lattice spacing artefacts
are revealed at small $r \lesssim 3\, a$.  The ${\cal O}(a^2)$ improvement
in the L\"{u}scher-Weisz Coulomb term is also readily apparent.

\begin{figure}[!t]
\null\hfill\includegraphics[angle=90,width=0.983\hsize]{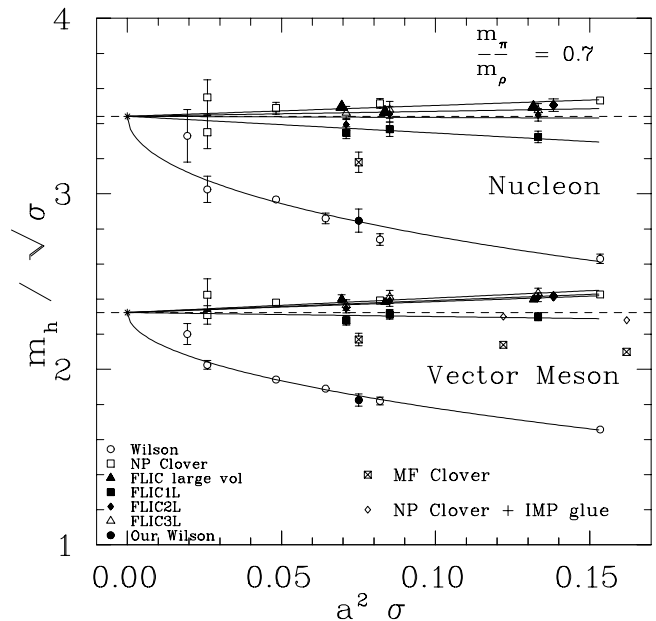}

\vspace{20pt}

\includegraphics[angle=90,width=\hsize]{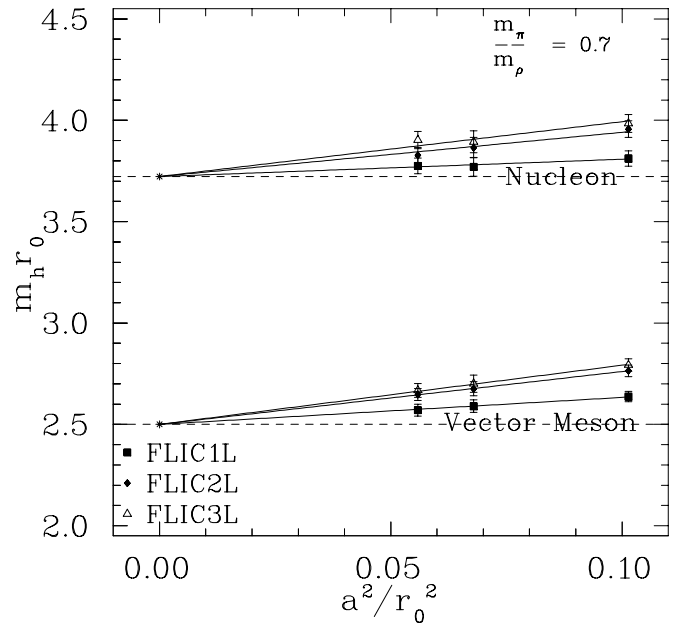}
\caption{The scaling of the $N$ and $\rho$ masses for various quark
  actions in the quenched approximation according to the string
  tension (upper) and the Sommer scale (lower). \label{fig:scale}} 
\end{figure}

\section{Scaling Results}
\label{scalingResults}

Calculations are performed on mean-field improved plaquette plus
rectangle $SU(3)$ L\"{u}scher-Weisz lattices.  Lattice spacings
determined using fits to Eq.~(\ref{eq:4fit}) above are given in
Table~\ref{tab:spacing}.

\begin{table}[h]
\begin{ruledtabular}
\begin{tabular}{ccc}
$\beta$ & $a[\sigma]{\rm (fm)}$ & $a[r_0]{\rm (fm)}$ \\
\hline
4.80 & 0.096(1) & 0.088(1) \\
4.60 & 0.120(1) & 0.113(1) \\
4.53 & 0.132(1) & 0.124(1) \\
4.38 & 0.164(1) & 0.152(1)
\end{tabular}
\end{ruledtabular}
\caption{The lattice spacing for pure L\"{u}scher-Weisz glue
  determined by the string tension $\sqrt{\sigma}= 440\text{ MeV}$ and
  the Sommer scale $r_0=0.49\text{ fm}$ for various couplings
  $\beta$. \label{tab:spacing}} 
\end{table}

\begin{table}[b]
\begin{ruledtabular}
\begin{tabular}{cccccc}
& $\beta$ & $M_N/\sqrt{\sigma}$ & $M_\rho/\sqrt{\sigma}$ & $M_N r_0 $ & $M_\rho r_0 $ \\
\hline
FLIC-1L & 4.60  & 2.278(26)  & 3.347(33)  & 2.638(30)  & 3.875(39) \\
& 4.53  & 2.313(27)  & 3.368(41)  & 2.662(31)  & 3.876(47) \\
& 4.38  & 2.299(21)  & 3.323(32)  & 2.688(25)  & 3.886(38) \\
\hline
FLIC-2L & 4.60  & 2.347(26)  & 3.394(33)  & 2.717(30)  & 3.929(39) \\
& 4.53  & 2.39(30)  & 3.453(44)  & 2.751(35)  & 3.974(51) \\
& 4.38  & 2.41(24)  & 3.450(35)  & 2.818(28)  & 4.034(41) \\
\hline
FLIC-3L & 4.60  & 2.365(30)  & 3.461(37)  & 2.738(34)  & 4.006(43) \\
& 4.53  & 2.413(37)  & 3.478(48)  & 2.776(43)  & 4.003(55) \\
& 4.38  & 2.435(27)  & 3.474(38)  & 2.847(32)  & 4.062(44)
\end{tabular}
\end{ruledtabular}
\caption{Results for the $N$ and $\rho$ masses on the three lattices,
  for the scale determined by the string tension $\sigma$ and the
  Sommer scale $r_0$. \label{tab:masses}} 
\end{table}

\begin{figure*}[!bt]
\includegraphics[angle=90,width=0.42\textwidth]{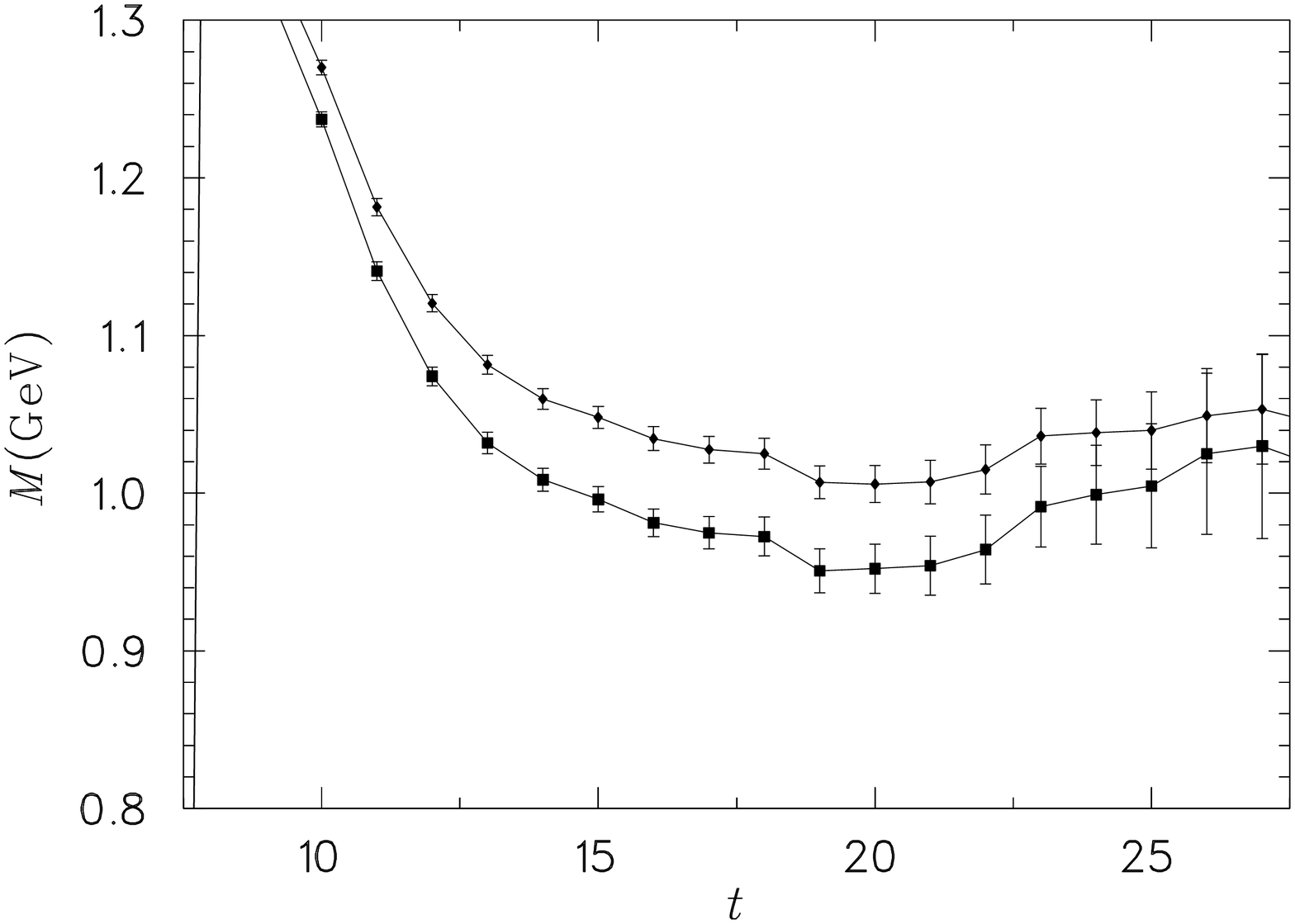}
\includegraphics[angle=90,width=0.42\textwidth]{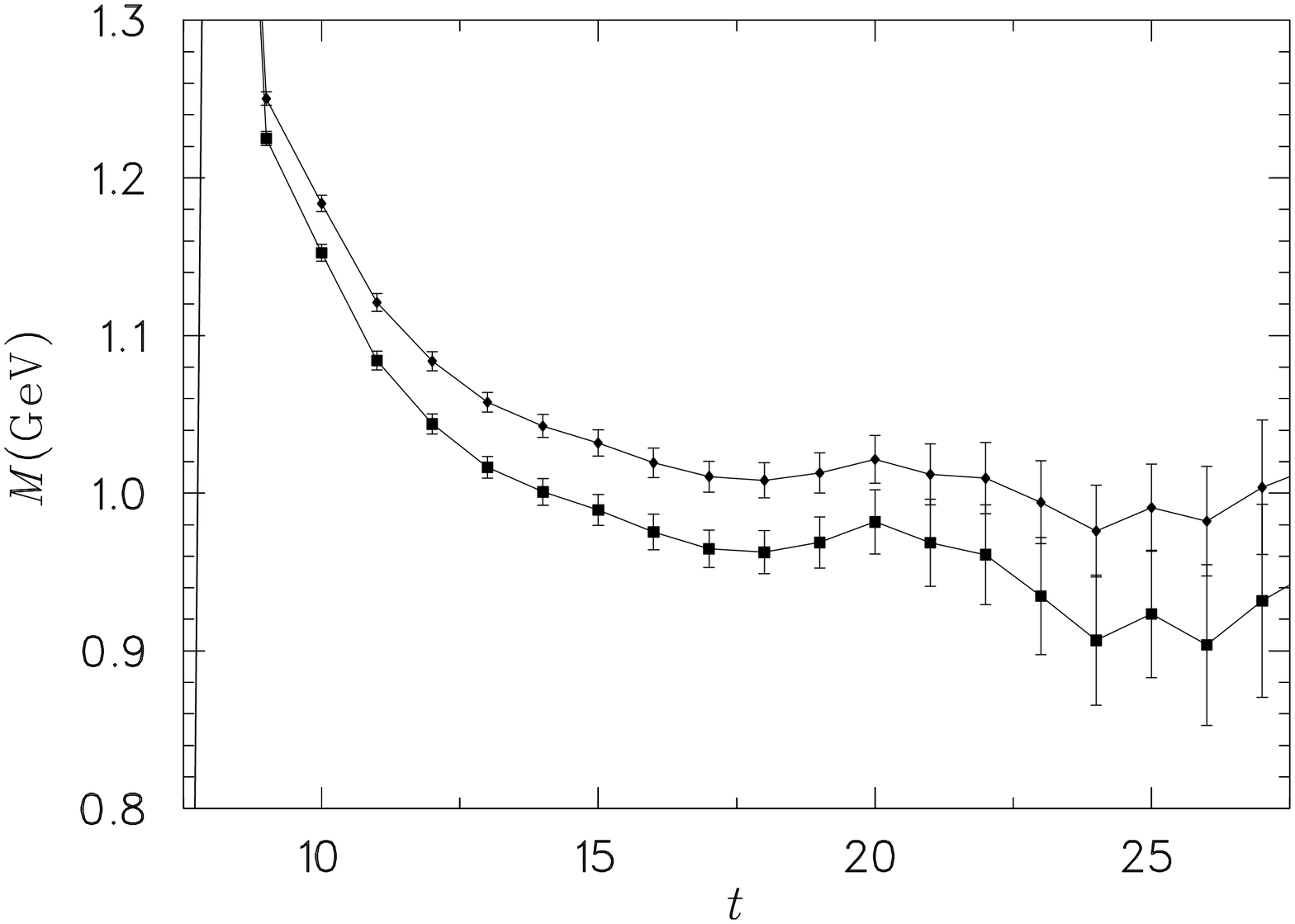}

\includegraphics[angle=90,width=0.42\textwidth]{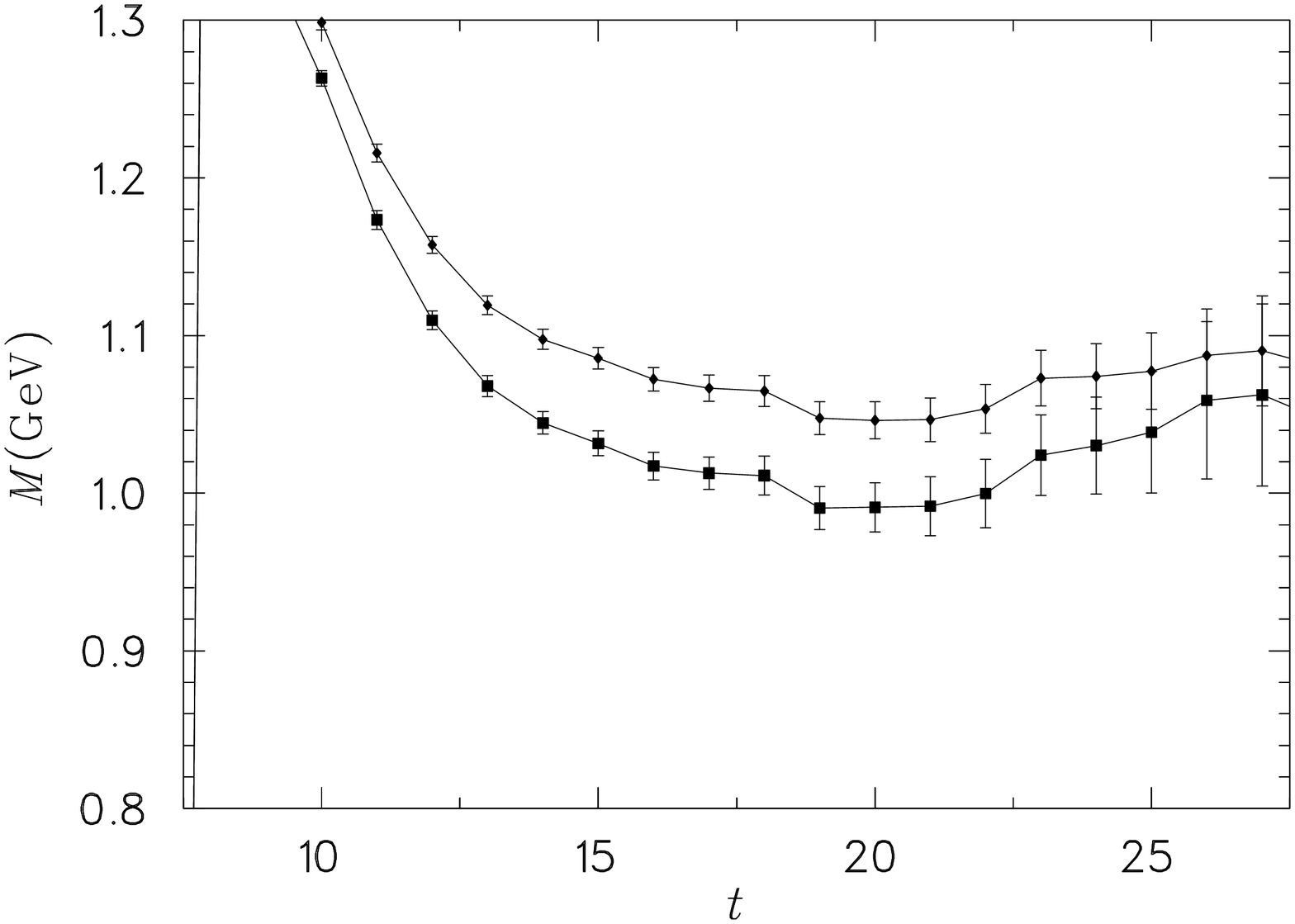}
\includegraphics[angle=90,width=0.42\textwidth]{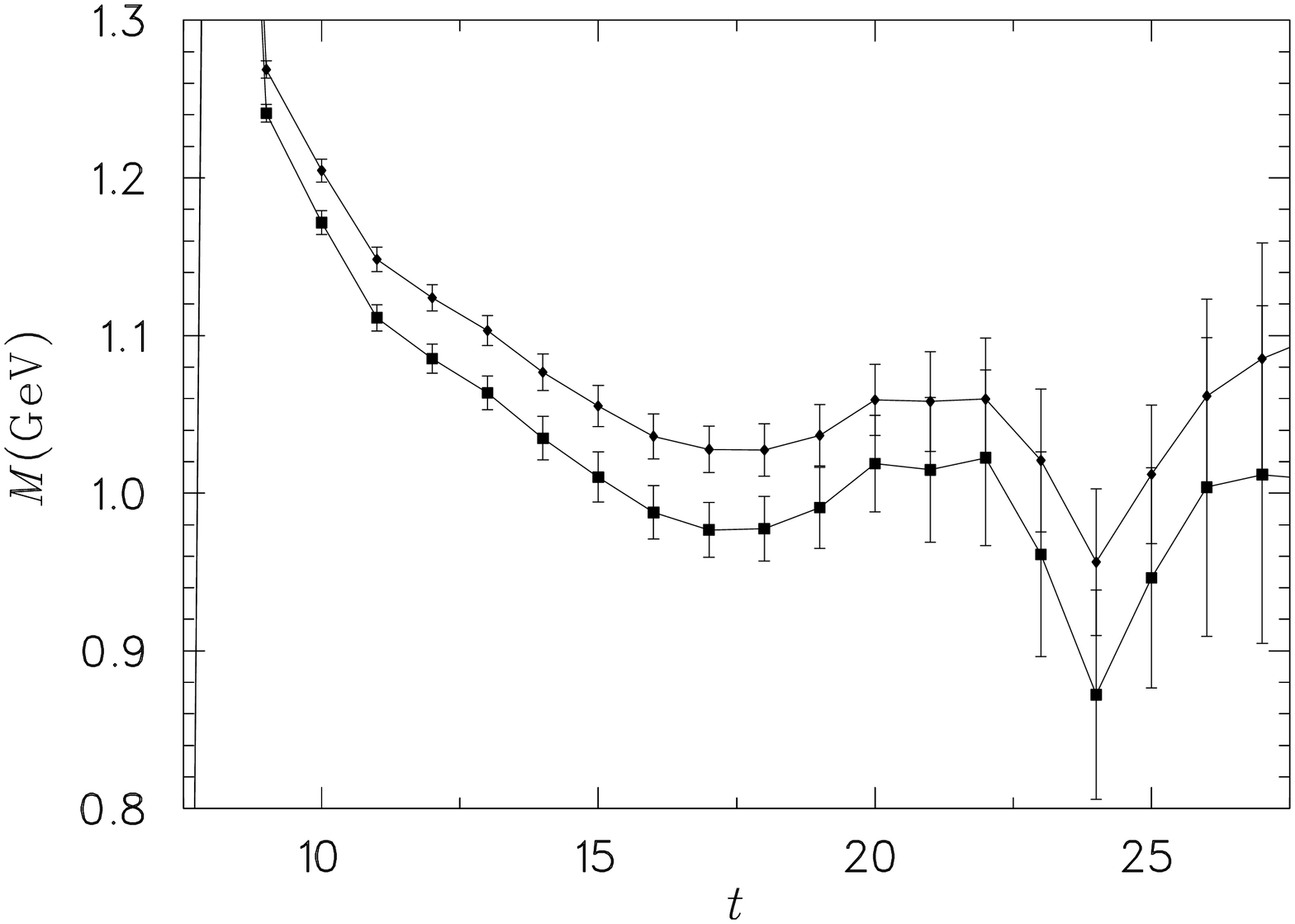}

\includegraphics[angle=90,width=0.42\textwidth]{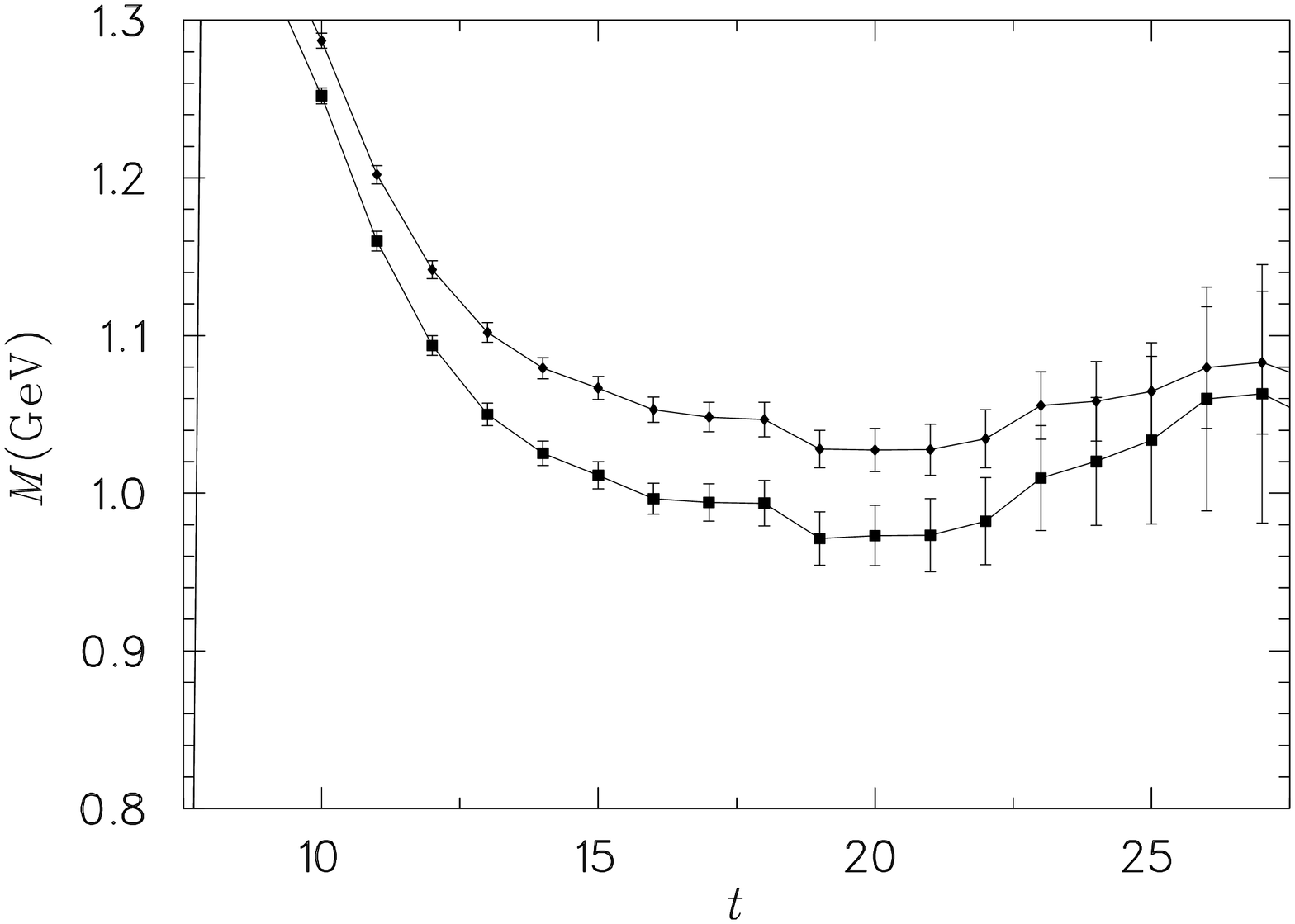}
\includegraphics[angle=90,width=0.42\textwidth]{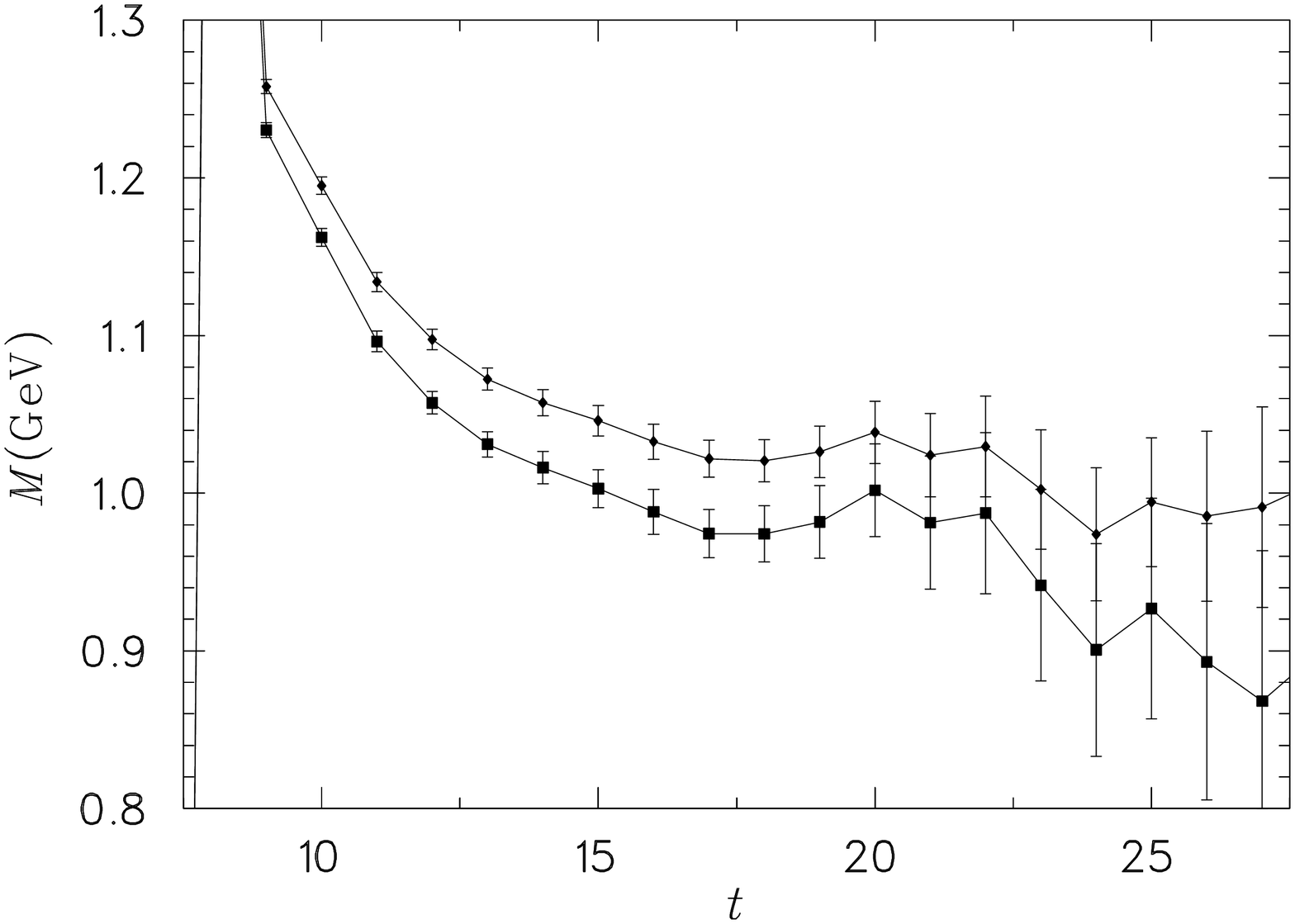}

\caption{$\rho-$meson correlation functions at two approximately
  matched quark masses for the 1-loop (top), 2-loop (middle) and
  3-loop (bottom) FLIC actions. Two lattices at $16^3\times 32$ are
  shown, $\beta=4.60$ (left) and $\beta = 4.53$
  (right). \label{fig:rhomass}} 
\end{figure*}

For each of the lattices we calculate quark propagators using the FLIC
fermion action with a 1, 2 and 3-loop clover term as described in
Sec.~\ref{FLICfermions}.  The $\pi,\rho$ and $N$ masses are then
calculated and interpolated to a $\pi/\rho$ mass ratio of 0.7, shown
in Table~\ref{tab:masses}.

Scaling results are presented in Fig.~\ref{fig:scale}.  The lines of fit
are extrapolations in $a^2$ constrained to pass through the single
point at the continuum limit.  The lines for non-perturbatively improved
clover and all FLIC actions are straight, indicating $O(a^2)$ scaling,
that is the effective elimination of $O(a)$ errors.

Thus, 1-, 2- and 3-loop fat-link formulations of $F_{\mu\nu}$ in the
FLIC fermion action all provide $O(a)$ improvement as expected.  The
different formulations differ at the level of $O(a^2).$ Remarkably,
the 1-loop action is actually the preferred action.  Firstly, it is
the cheapest to perform molecular dynamics with, which is important
for Hybrid Monte Carlo dynamical simulations.  Secondly it has the
smallest residual $O(a^2)$ errors in the quantities we have studied
here.  We'll also see that correlation functions have smaller
fluctuations. 

\section{Correlation Functions}
\label{correlationFunctions}

Finally, we compare the $\rho-$meson correlation function on the fine
$\beta=4.60$ and coarse $\beta=4.53$ lattices at approximately matched
pion masses for the three different FLIC actions.  The source is at
time slice 8. 

The effective mass plots are given in Figure~\ref{fig:rhomass}.  The
main effect that we observe is that as the Euclidean time index
progresses into the latter half of the lattice, the 1-loop FLIC
correlators show reduced fluctuations and reduced error bars when
compared with the 2-loop and 3-loop FLIC results.  The difference is
particularly observable on the coarser $\beta=4.53$ lattice.  We
understand this to be due to the 1-loop action having a more local
field strength $F_{\mu\nu}(x)$ than the 2- and 3-loop actions making
it less susceptible to large fluctuations.

\section{Conclusions}
\label{conclusions}

We have examined the role of improvement in the lattice field strength
tensor of the FLIC fermion action, Our results demonstrate that the
standard 1-loop choice of for the lattice clover term in the FLIC
fermion action provides $O(a^2)$ scaling.  

Remarkably the 1-loop action provides results that are preferable to
those obtained from the 2-loop ${\cal O}(a^2)$-improved lattice field
strength tensor or those obtained from the the 3-loop ${\cal
O}(a^4)$-improved definition.  The 1-loop results provide
\begin{enumerate}
\item Smaller residual $O(a^2)$ errors,
\item Stable hadron correlators with reduced fluctuations,
\item Smaller statistical uncertainties, and 
\item A more efficient action suitable for dynamical fermion
  simulations. 
\end{enumerate}

This result enables efficient and effective dynamical QCD simulations
with FLIC fermions.  Simulations are currently under way.

\section*{Acknowledgments}

We thank the Australian Partnership for Advanced Computing (APAC) and
the South Australian Partnership for Advanced Computing (SAPAC) for
generous grants of supercomputer time which have enabled this project.
This work is supported by the Australian Research Council.


\end{document}